\documentclass[conference]{IEEEtran}
\pdfoutput=1
\usepackage[utf8]{inputenc}

\usepackage{graphicx}
\usepackage{textcomp}
\usepackage[dvipsnames]{xcolor}
\usepackage{cite}
\usepackage{enumitem}
\usepackage{verbatim}

\usepackage{amsmath,amssymb,amsfonts}
\usepackage[ruled,vlined]{algorithm2e}

\usepackage[normalem]{ulem}
\usepackage{subcaption}

\SetCommentSty{mycommfont}

\usepackage{tikz}
\usepackage{pgfplots, pgfplotstable}

\usepackage[textsize=tiny,textwidth=1.4cm,disable]{todonotes}
\newcommand{\Aron}[1]{\todo[color=yellow!5,linecolor=black!50]{\textbf{Aron}: #1}}

\newcommand{\Michael}[1]{\todo[color=orange!5,linecolor=black!50]{\textbf{Michael}: #1}}
\newcommand{\ad}[1]{\todo[color=green!5,linecolor=black!50]{\textbf{Abhishek}: #1}}

\pgfplotsset{
SmallBarPlot/.style={
    font=\footnotesize,
    ybar,
    width=\linewidth,
    ymin=0,
    xtick=data,
    xticklabel style={text width=0.8cm, align=center},
    xtick pos=left,
},
BlueBars/.style={
    fill=MidnightBlue!30, bar width=1.5
},
RedBars/.style={
    fill=red!40, bar width=2
}
}
\pgfplotsset{grid style={dashed,gray}}
\pgfplotsset{minor grid style={dashed,red}}
\pgfplotsset{major grid style={dotted,green!50!black}}
\pgfplotsset{compat=1.6}
\usetikzlibrary{scopes}
\usetikzlibrary{calc,chains,positioning,shapes.geometric,shadows}

\usepackage{url}
\usepackage[capitalise,noabbrev,nameinlink]{cleveref}

\begin{document}

% workaround for todonotes
% \clearpage
% \setcounter{page}{1}

\setlength{\marginparwidth}{1.38cm}

\title{Data-Driven Prediction of Route-Level Energy Use for Mixed-Vehicle Transit Fleets}

\author{\IEEEauthorblockN{Afiya Ayman\IEEEauthorrefmark{1}, Michael Wilbur\IEEEauthorrefmark{2}, Amutheezan Sivagnanam\IEEEauthorrefmark{1}, Philip Pugliese\IEEEauthorrefmark{3}, Abhishek Dubey\IEEEauthorrefmark{2}, Aron Laszka\IEEEauthorrefmark{1}}
\IEEEauthorblockA{\IEEEauthorrefmark{1}\textit{University of Houston}, \IEEEauthorrefmark{2}\textit{Vanderbilt University}, \IEEEauthorrefmark{3}\textit{Chattanooga Area Regional Transportation Authority}}
}

% REMOVE BEFORE SUBMISSION
%\pagestyle{plain}

\maketitle

\begin{center}
  Published in the proceedings of the\\6th IEEE International Conference on Smart Computing (SMARTCOMP 2020).
\end{center}

\begin{abstract}
Due to increasing concerns about environmental impact, operating costs, and energy security, public transit agencies are seeking to reduce their fuel use by employing electric vehicles (EVs).
However, because of the high upfront cost of EVs, most agencies can afford only mixed fleets of internal-combustion and electric vehicles.
Making the best use of these mixed fleets presents a challenge for agencies since optimizing the assignment of vehicles to transit routes, scheduling charging, etc.\ require accurate predictions of electricity and fuel use. 
Recent advances in sensor-based technologies, data analytics, and machine learning enable remedying this situation; however, to the best of our knowledge, there exists no framework that would integrate all relevant data into a route-level prediction model for public transit. 
In this paper, we present a novel framework for the data-driven prediction of route-level energy use for mixed-vehicle transit fleets, which we evaluate using data collected from the bus fleet of CARTA, the public transit authority of Chattanooga, TN.
We present a data collection and storage framework, which we use to capture system-level data, including traffic and weather conditions, and high-frequency vehicle-level data, including location traces, fuel or electricity use, etc.
We present domain-specific methods and algorithms for integrating and cleansing data from various sources, including street and elevation maps.
Finally, we train and evaluate machine learning models, including deep neural networks, decision trees, and linear regression, on our integrated dataset.
Our results show that neural networks provide accurate estimates, while other models can help us discover relations between energy use and factors such as road and weather conditions.
\end{abstract}

\section{Introduction}
\label{sec:intro}

% ENVIRONMENTAL IMPACT
Transportation accounts for 28\% of the total energy use in the U.S.~\cite{eia}, and as such, it is responsible for immense environmental impact, including urban air pollution and greenhouse gas emissions, and may pose a severe threat to energy security. 
Switching from personal vehicles to public transit systems can significantly reduce energy use and environmental impact.
However, even public transit systems require substantial amounts of energy; for example, public bus transit services in the U.S. are responsible for at least 19.7 million metric tons of CO$_2$ emission annually~\cite{EPA-420-F-19-047}.

% ELECTRIC VEHICLES AND PLANNING
Electric vehicles (EVs) can have much lower environmental impact during operation than comparable internal combustion engine vehicles (ICEVs), especially in urban areas.
However, existing EVs have limited battery capacity and hence driving range.
For example, a BYD K9S bus has a nominal driving range of only around 150 miles\Aron{according to Wikipedia}.
Due to this limited driving range, the operation of EVs must be carefully planned.
% TRANSIT AGENCY PLANNING
Such planning is especially important to transit agencies that operate mixed fleets of electric and internal-combustion vehicles.
Firstly, these agencies need to decide which vehicles are assigned to serving which transit trips.
Since the advantage of EVs over ICEVs varies depending on the route and time of day (e.g., the advantage of EVs is higher in slower traffic with frequent stops, and lower on highways), the assignment can have a significant effect on energy use and, hence, environmental impact.  
Secondly, they need to schedule when to charge electric vehicles during the day considering how long EVs can operate without recharging and when electricity prices are lower. 
%Because transit agencies often have limited charging capabilities (e.g., limited number of charging poles, or limited maximum power to avoid high peak  loads on the electric grid), charging constraints can significantly increase the complexity of the assignment and scheduling problem. 
%Unfortunately, EVs are also much more expensive than ICEVs (typically, diesel transit buses cost less than \$500K, while electric ones cost more than \$700K, or close to around \$1M with charging infrastructure).

% PREDICTION
At the crux of this operational optimization is 
the problem of \emph{accurately predicting the electricity and fuel consumption} of transit vehicles.
Such predictions must be contextualized with a variety of factors, including the type of vehicle, traffic and weather conditions, road gradient, and type of road (e.g., highway vs. residential area) since these factors can have significant impact on energy use.
Clearly, handling all of these factors using model-driven approaches, which attempt to build detailed physical models of vehicles, is very challenging.

% CHALLENGES
Recent advances in sensor-based technologies, data analytics, and machine learning have enabled remedying this situation by building data-driven predictors of route-level energy use.
However, to the best of our knowledge, there exists no framework that would integrate all relevant data into a route-level prediction model for public transit. 
Such a framework needs to address many challenges:
high volume of unstructured and irregular data must be stored efficiently, allowing easy retrieval in subsequent steps;
noisy data (e.g., GPS based locations) must first be cleansed (e.g., corrected or imputed based on other data sources);
heterogeneous data (recorded at different rates with different precision in different formats) must be collated into samples that can be fed into training machine-learning models;
etc.

% CONTRIBUTIONS
\textbf{Contributions:}
In this paper, we present a novel framework for the data-driven offline prediction of route-level energy use for mixed-vehicle transit fleets, which we evaluate using data collected from the bus fleet of CARTA, the public transit authority of Chattanooga, TN.
\begin{itemize}[topsep=0pt, leftmargin=*]
    \item We collect and combine vehicle telemetry data, elevation and street-level maps, weather data, and traffic data.
    Our dataset is publicly available at \texttt{\url{https://hdemma.github.io/}}\Aron{Make sure that it is!}
    \item We present a cloud-centric data collection and storage framework for high-velocity spatiotemporal smart-city data. Our modular architecture is centered around a topic-based distributed log with easily extendable, application-specific structured views.
    %\item We present a data storage and view architecture for high-velocity data.
    \item We present a framework and novel algorithms for cleaning and integrating time series data from multiple sources into sets of samples with fixed-dimension feature space.
    \item We train machine-learning models on this dataset (deep neural networks, linear regression, and decisions trees) and study their performance, focusing on the impact of including or omitting certain data sources.
\end{itemize}

\textbf{Organization:}
The remainder of this paper is organized as follows.
\Aron{Check and update right before submission (maybe merge into previous paragraph?)}
In \cref{sec:dataCollection}, we describe our data sources, data collection methods, and data storage architecture.
In \cref{sec:dataProcessing}, we introduce our data cleansing and integration framework.
In \cref{sec:predictionModels}, we propose machine-learning based prediction models.
In \cref{sec:numerical}, we present numerical results based on real-world data.
In \cref{sec:related}, we discuss related work.
Finally, in \cref{sec:concl}, we provide concluding remarks.

\section{Data Collection and Storage}
\label{sec:dataCollection}

\Aron{remove to make space?}We first provide an overview of the data sources that we use in our study (\cref{sec:datasources}) and then describe the architecture of our data storage framework (\cref{sec:dataarchitecture}).

\subsection{Data Sources}
\label{sec:datasources}

\subsubsection{Vehicle Data}

\begin{table}[t]
\centering
\caption{Overview of Vehicle Dataset}
\label{tab:Dataset_Description}
\resizebox{\linewidth}{!}{
\renewcommand{\arraystretch}{1.2}
\setlength{\tabcolsep}{2.5pt}
% \begin{tabular}{|c||c|c|c|c|c|c|}
% \hline
% \multicolumn{1}{|c||}{\begin{tabular}[c]{@{}c@{}}\textbf{Vehicle}\\ \textbf{Type}\end{tabular}}
% & \multicolumn{1}{|c|}{\begin{tabular}[c]{@{}c@{}}\textbf{CARTA}\\ \textbf{Vehicle ID}\end{tabular}}
% & \textbf{Model} & \multicolumn{1}{c|}{\begin{tabular}[c]{@{}c@{}}\textbf{Model}\\ \textbf{Year}\end{tabular}} & \textbf{Start Date}&\textbf{End Date}&\textbf{Duration}\\
% % & \textbf{Train Set}& \textbf{Test Set}\\
% \hline
% \hline 
% \textcolor{blue}{Diesel} & \multicolumn{1}{|c|}{\begin{tabular}[c]{@{}c@{}}Gillig\\ 148, 149, 150\end{tabular}} & {\begin{tabular}[c]{@{}c@{}}Gillig Phantom\\Diesel Allison Cummins\end{tabular}} & 2014 &2019-08-22 & 2019-10-16 & 56 days\\%& 85\% & 15\% \\
% \hline
% \hline 
% \textcolor{red}{Electric} & \multicolumn{1}{|c|}{\begin{tabular}[c]{@{}c@{}}BYD\\ 751, 752, 753\end{tabular}} & {\begin{tabular}[c]{@{}c@{}}BYD K9S\\35 feet Battery-Electric\end{tabular}} & 2016 &2019-08-01 & 2019-10-01 & 61 days\\%& 85\% & 15\% \\
% \hline
% \end{tabular}
\begin{tabular}{|c||c|c|c|c|c|}
\hline
\multicolumn{1}{|c||}{\begin{tabular}[c]{@{}c@{}}\textbf{Vehicle}\\ \textbf{Type}\end{tabular}}
& \multicolumn{1}{|c|}{\begin{tabular}[c]{@{}c@{}}\textbf{CARTA}\\ \textbf{Vehicle ID}\end{tabular}}
& \textbf{Model}  & \textbf{Start Date}&\textbf{End Date}&\textbf{Duration}\\
% & \textbf{Train Set}& \textbf{Test Set}\\
\hline
\hline 
\textcolor{blue}{Diesel} & \multicolumn{1}{|c|}{\begin{tabular}[c]{@{}c@{}}Gillig\\ 148, 149, 150\end{tabular}} & {\begin{tabular}[c]{@{}c@{}}2014 Gillig Phantom\\diesel Cummins-Allison\end{tabular}} & 2019-08-22 & 2019-10-16 & 56 days\\%& 85\% & 15\% \\
\hline
\hline 
\textcolor{red}{Electric} & \multicolumn{1}{|c|}{\begin{tabular}[c]{@{}c@{}}BYD\\ 751, 752, 753\end{tabular}} & {\begin{tabular}[c]{@{}c@{}}2016 BYD K9S\\35-foot battery-electric\end{tabular}} & 2019-08-01 & 2019-10-01 & 61 days\\%& 85\% & 15\% \\
\hline
\end{tabular}
}
\end{table}

To collect data from CARTA's fleet of vehicles, we partner with ViriCiti, a company that offers sensor devices and an online platform to support transit operators with real-time insight into their fleets. ViriCiti has installed sensors on CARTA's mixed-fleet of 3 electric, 41 diesel, and 6 hybrid buses, and it has been collecting data continuously %for over a year 
at 1-second (or shorter) intervals since installation.
%This dataset includes location traces from GPS, real-time fuel and electricity use, battery charge, etc.
At the time of this study, we have 2 months of data available for 3 electric and 3 diesel buses, on which sensors were installed earliest (see \cref{tab:Dataset_Description}).
%For building our data-driven prediction model, we select two months of data from 3 electric and 3 diesel buses. 
All electric buses are BYD K9S %35-foot
battery-electric transit vehicles, while the diesel buses are 2014 Gillig Phantom series vehicles with Cummins diesel engines. % and Allison transmissions.

For each vehicle, we obtain time series data from ViriCiti, which includes series of timestamps and vehicle locations based on GPS. For electric buses, we also include features such as battery current in ampere ($A$), battery voltage ($V$), battery state of charge, and charging cable status. For diesel buses, we include fuel level and the total amount of fuel used over time in gallons.
In total, we have already obtained around 6.6 million data points for electric buses and 1.1 million data points for diesel buses (\cref{tab:Dataset_Description}). Fuel data is recorded less frequently; hence, there are fewer data points for diesel~buses.

\subsubsection{Elevation, Weather, and Traffic Data} \label{Subsection: Elevation, Weather and Traffic Data}

We collect static GIS elevation data from the Tennessee Geographic Information Council~\cite{tnelevationdata}. From this source, we download high-resolution digital elevation models (DEMs), derived from LIDAR elevation imaging, with a vertical accuracy of approximately 10 cm \cite{usgselevationdata}. We join the DEMs for Chattanooga into a single DEM file, which we  then use to determine the elevation of any location within the geographical region of our~study. \ad{guys we need to acknowledge Brian Xu for his help with this}

We collect weather data from multiple weather stations in Chattanooga at 5-minute intervals using the DarkSky API~\cite{darkskyapi}. This data includes real-time temperature, humidity, air pressure, wind speed, wind direction, and precipitation.

We collect traffic data at 1-minute intervals using the HERE API~\cite{hereapi},
which provides speed recordings for segments of major roads. %, which provides data in the form of timestamped speed recordings from selected roads.
Every road segment is identified by a unique Traffic Message Channel identifier (TMC ID) \cite{tmc}. Each TMC ID is also associated with a list of latitude and longitude coordinates, which describe the geometry of the road segment. Weather and traffic data was collected from August 1, 2019 to October 1, 2019 to match the time range in \cref{tab:Dataset_Description}.

\subsection{Data Architecture Framework}
\label{sec:dataarchitecture}

%In this section we present our cloud-centric data storage and processing framework. The overview of our framework is provided in figure \ref{fig:data_architecture_overview} while the implementation is presented in figure \ref{fig:data_architecture_implementation}.

%\subsubsection{Challenges and Design}

% \begin{figure}[!t]
% \begin{center}
% \centerline{\includegraphics[width=1.0\columnwidth]{figures/data_arch_overview.png}}
% \caption{Data Architecture Overview}
% \label{fig:data_architecture_overview}
% \end{center}
% \vspace{-0.2in}
% \end{figure}

\begin{figure}[t]
\centering
\begin{tikzpicture}
[
    >=stealth,
    semithick,
    % node distance=3.5cm,
    % node distance = 1.5cm, 
    alg/.style = {draw, align=center, font=\linespread{0.95}\selectfont\footnotesize, rounded corners=0.15em, drop shadow=gray, inner sep=0.6em}
  ]
 \tikzset{node distance = 1.4cm} 
    \begin{scope}[node distance=3.25cm]
        \node (n2) [alg,fill=blue!10] {Real-time\\Data};
        \node (n3) [alg,right of=n2,fill=orange!14]  {Pub-Sub \\ (Distributed Ledgers)};
        \node (n4) [alg,right of=n3,fill=orange!14]  {Structured\\Views (NoSQL)};
    \end{scope}
    \node (n1) [alg,above of=n4,fill=blue!10] {Static Data};
    \node (n7) [alg,below of= n2,fill=blue!10]  {Client\\Application};
    
    \begin{scope}[node distance=3.25cm]
    
    \node (n6) [alg,right of=n7,fill=red!14]  {Real-Time Model\\Inference};
    \node (n5) [alg,right of=n6,fill=red!14] {Model\\Training};
    \end{scope}

    \draw[->,black] (n2) --  node[black,midway,right,sloped,font=\scriptsize]{} (n3);
    \draw[->,black] (n3) --  node[black,midway,right,sloped,font=\scriptsize]{} (n4);
    \draw[->,black] (n4) --  node[black,midway,below,sloped,font=\scriptsize]{} (n5);
    \draw[->,black] (n5) --  node[black,midway,left,sloped,font=\scriptsize]{} (n6);
    \draw[->,black] (n6) --  node[black,midway,left,sloped,font=\scriptsize]{} (n7);
    \draw[->,black] (n1) --  node[black,midway,below,sloped,font=\scriptsize]{} (n4);
    \draw[->,black] (n3) --  node[black,midway,below,sloped,font=\scriptsize]{} (n6);
    
    % \draw[->,blue!50] (db1) -- ++(0,1) -- ($(db2)+(0,1)$) node[black,midway,above,font=\scriptsize]{Link: Name} node[black, midway,below,font=\scriptsize]{Owner: Name} -- (db2) ;
  \end{tikzpicture}
  \caption{Data architecture overview.}
  \vspace{-0.1in}
\label{fig:data_architecture_overview}
    
\end{figure}
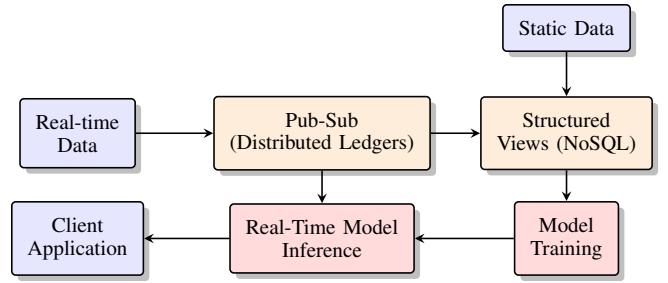

Next, we outline a general-purpose data architecture framework for storing the various smart-city data streams. The goal of this framework is to store the data streams in a way that provides easy access for offline model training and updates as well as real-time access for system monitoring and prediction. An overview of our architecture is shown in \cref{fig:data_architecture_overview}.

The first challenge is persistent storage of the high-velocity, high-volume data streams. In this study, the real-time data sources---ViriCiti, HERE, and DarkSky---produce around 100 GiB of data per month. Therefore, we choose a cloud based design to allow for fast horizontal scalability of the system. 

The second concern is that the data itself is highly unstructured and irregular. Additionally, each data source streams at varying rates. Therefore we stream each data source to a topic-based publish-subscribe (pub-sub) layer which persistently stores each data stream as a separate topic. All replication is handled at the ledger level, which allows downstream storage and applications to adapt and expand without concern for data resiliency. The distributed ledgers are append only logs and store incoming data in its raw, unstructured form. This data structure allows for near real-time access to incoming data, which is optimal for model inference during deployment and latency-sensitive client applications such as monitoring or visualization tools. This setup minimizes latency for running trained models in production in real-time use cases. Data streams are accessed by unique topic names, and data is persisted in each ledger, allowing for historical~access. 

Model-training and inference require data from various streams to be merged. Typical implementations of stream processing architectures require external processing frameworks such as Apache Spark and Storm \cite{zaharia2016apache,iqbal2015big}. For our system we instead incorporate a customized stream processing layer into the pub-sub module. In this layer, data cleansing and processing functions are applied to the raw data topics and the processed data is then published to separate reformed topics that can easily be accessed for prediction or model training.

\begin{figure}[!t]
\centering
\begin{tikzpicture}[
  x=2.25cm,
  thick,
  font=\scriptsize\bfseries,
  Node/.style={align=left, rounded corners=0.15em},
  DataNode/.style={Node, text=white, fill=BrickRed, inner sep=0.16cm, minimum width=1.2cm},
]
\draw [orange, rounded corners=0.15em] (0.5, -1.99) rectangle (2.45, 1.2);
\draw [MidnightBlue, rounded corners=0.15em] (0.45, -2.1) rectangle (3.6, 1.5);
\node [DataNode] (viriciti) at (0, -0.25) {ViriCiti};
\node [DataNode, above of=viriciti] {Traffic};
\node [DataNode, below of=viriciti] {Weather};
\node [Node, font=\scriptsize] (brokers) at (1, -1.05) {Brokers \& Bookies\\ (5 nodes)\\ \textbullet~2\,VPUs\\ \textbullet~8\,GB RAM\\ \textbullet~40\,GiB root disk\\ \textbullet~128\,GiB volume};
\node [above=-0.1cm of brokers] {\includegraphics[width=0.75cm]{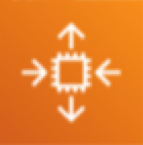}};
\node [Node, font=\scriptsize] (zookeeper) at (2, -1.05) {Zookeeper\\ (5 nodes)\\ \textbullet~1\,VPU\\ \textbullet~8\,GB RAM\\ \textbullet~40\,GiB root disk\\ \textbullet~64\,GiB volume};
\node [above=-0.1cm of zookeeper] {\includegraphics[width=0.75cm]{figures/node.png}};
\node [DataNode, fill=Orange!85!black] at (1.5, 1.15) {Pulsar Cluster};
\node [Node, font=\scriptsize] (mongodb) at (3.1, -0.9) {MongoDB\\ \textbullet~8\,VPUs\\ \textbullet~16\,GB RAM\\ \textbullet~150\,GiB root disk\\ \textbullet~4\,TB volume};
\node [above=-0.1cm of mongodb] {\includegraphics[width=1.2cm]{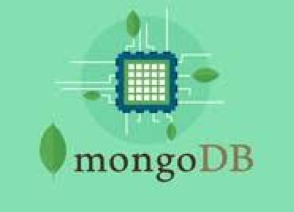}};
\node [DataNode, fill=MidnightBlue] at (3.1, 1.4) {Cloud};
\draw [->] (0.3, -0.25) -- (0.4, -0.25);
\draw [->] (2.5, -0.25) -- (2.6, -0.25);
\end{tikzpicture}
\caption{Data architecture implementation.}
\vspace{-0.15in}
\label{fig:data_architecture_implementation}
\end{figure}

As shown in \cref{fig:data_architecture_implementation}, we use the pub-sub framework Apache Pulsar \cite{apachepulsar} for the topic-based distributed ledger module. Apache Pulsar provides topic-based messaging. The storage component of Apache Pulsar relies on Apache BookKeeper \cite{apachebookkeer}, which allows sharding of data at the topic level. As the size and velocity of data varies greatly between data sources, topic level sharding allowed us to evenly distribute data between storage nodes and thus maximize resources in the cluster. Cluster state and coordination is managed with Apache ZooKeeper \cite{apachezookeeper}. The Apache Pulsar system provides automatic failover and load balancing.

While distributed topic-based ledgers provide fast real-time access to data and easy data replication, the complexity of working with spatiotemporal data requires a more structured representation of the data, particularly for training and batch analysis. Therefore, we incorporate a structured view component into the architecture downstream from the distributed ledgers. In this sense, structured views are data representations optimized for specific downstream components. For our use case, this includes model training and data analysis client applications. These applications require a data model with spatial and temporal indexing for efficient data retrieval, which is particularly important during model training. Additionally, large-scale data has to be shared between research sites, which requires a unified structure that is easily transferable. For this, we use MongoDB \cite{mongodb}, which provides native geospatial indexing and easy large-scale exports in JSON format for sharing between research sites. \ad{mike can you produce some results that show why this architecture is better than a simple replicated SQL architecture.} \Michael{I do not have any results on this. We can set up some type of simulation to compare which can be part of an extension}

%\subsubsection{Implementation}

\section{Data Processing Framework}
\label{sec:dataProcessing}

Before applying machine-learning models, we have to process the time series data recorded from the vehicles by cleaning it, generating samples with a fixed-dimension feature space, and integrating with other data sources.

%Before training our predictors, we process the collected data in multiple steps. Here, we describe each step of this data processing workflow in detail, starting with cleaning the vehicle data obtained from ViriCiti. ViriCiti provides real-time vehicle data in the form of a time series of tuples, which include features such as timestamps and GPS-based locations. 

\subsection{Removing Garage Locations and Charging}

Since our goal is to predict the amount of energy used for driving, we remove all datapoints that were recorded when a bus was (1) waiting in the garage or (2) charging. 
%When a bus is idle in the garage, it does not consume any energy. 
First, we remove all datapoints whose GPS-based locations fall in the geographical area of the CARTA bus garage. 
%Every time a vehicle leaves the garage, we consider that datapoint as the starting point for tracking energy consumption. % for the trip that the vehicle is about to take.
Second, for electric buses, we remove all datapoints whose charging cable status indicates that the vehicle was charging. % We also remove all records for electric buses where the bus is charging based on the charging cable status. 
%\Aron{We can omit this sentence since it is minor implementation detail.} The value of this feature is 1 when the vehicle is plugged into a charging station. \Aron{We can omit this sentence since it should be obvious to the reader.}This is done only for electric vehicles.

\subsection{Estimating Energy Use for Electric Vehicles}

For diesel buses, we can compute the amount of fuel used between two consecutive datapoints as the change in the total amount fuel used.
For electric buses, we could compute the amount of energy used as the change in the battery state of charge ($SoC$), which is the remaining battery charge as a percentage of the total capacity.
However, $SoC$ values are recorded with a low precision of only one digit after the decimal point.
To obtain more accurate values, we need to estimate the amount of energy used based on the recorded battery current ($A$) and voltage ($V$) values.
At any time, the instantaneous power use of the vehicle (in Watt) can be computed as $A \cdot V$.
We can estimate the amount of energy used (in Joule) between consecutive datapoints $i-1$ and $i$ as
\begin{equation}
    A_{i} \cdot V_{i} \cdot \left( TS_i - TS_{i-1}\right) , 
\end{equation}
where $TS_i$ is the timestamp of datapoint $i$ (in seconds).
Since current and voltage values are recorded at least once every second, the above formula provides a high-accuracy estimate. 
We confirmed that our estimates are unbiased by comparing them to changes in $SoC$ over large numbers of datapoints.

\begin{comment}
, and state of charge ($SoC$) values provided by ViriCiti. $SoC$ values are given as the remaining percentage of charge in the battery.  So, we convert the $SoC$ values to the amount of energy consumed in Joule ($J$) based on the total capacity of the battery of the electric vehicles, which is around
$1.8\cdot10^9$ Joules, and then recompute the $SoC$ for high precision. There are two cases: 

\subsubsection{When the bus leaves the garage} 
For a recorded time instant $i$ when the bus leaves the garage, we use the $SoC_{[i]}$ value recorded by ViriCiti.
    
\subsubsection{When the bus is moving}
At any time instant $i$, the energy consumed since the last recorded instant $i-1$ is power $P_{[i]} = A_{[i]} \cdot V_{[i]}$ multiplied by the time difference between the two records $TS_{[i]}$ and $TS_{[i-1]}$.
\begin{align*}
\sum Energy_{[i]} &= \sum P \cdot Time\textrm{ }Difference\\
&=\sum ( A_{[i]} \cdot V_{[i]}) \cdot \left(TS_{[i]} - TS_{[i-1]}\right)
\end{align*}

    % $$P = A_{[i]} \times V_{[i]}$$
    % $$Time\textrm{ }Difference = \frac{TS_{[i]} - TS_{[i-1]}}{1000}$$
    % $$\sum Energy_{[i]} = \sum P \times Time\textrm{ }Difference$$
    % $$ = \sum ( A_{[i]} \times V_{[i]}) \times \frac{TS_{[i]} - TS_{[i-1]}}{1000}$$
    
To transform the $\sum Energy_{[i]}$ into state of charge, % so that it is independent of any unit (second, joule or anything), 
we convert the $\sum Energy_{[i]}$ to a percentage value.
 
$$Computed\textrm{ }SoC_{[i]} = \frac{Total\textrm{ }Capacity - \sum Energy_{[i]}}{Total\textrm{ }Capacity} \cdot 100\%$$
\end{comment}

\subsection{Mapping GPS Locations to Roads}
\label{subsec:dataProcessing_mapping_GPS}

The recorded vehicle locations are inherently noisy since they are based on GPS. For example, some locations fall onto streets or parking lots where a bus cannot even drive. 
This noise presents a significant challenge for computing accurate travel distances and for integrating the time series with other data sources.
To mitigate this noise, we combine the recorded vehicle locations with %real street addresses using 
a street-level map of Chattanooga, which we obtain from OpenStreetMap (OSM). OSM represents each road using a disjoint set of segments, called OSM features. Specifically, OSM divides each road into one or more segments along its length and assigns a unique \emph{OSM Feature ID} to each one of these road~segments. 

We map each recorded GPS location to an OSM feature (i.e., road segment). For a particular location, we consider the set of nearby OSM features based on geographical distance.
For each nearby OSM feature, we count how many of the preceding and following datapoints were also near this feature.
Finally, we select the feature that is near the most datapoints.
\cref{algo:mapping} details the process of mapping locations to OSM~features.

% and compare them with the roads returned by a few of the prior and following coordinates. 
% For each of the road segment in the list of nearby streets, we look at the frequency in the results for the preceding and following \Aron{introduce constant}15 GPS locations and choose the most frequent road segment.
% \cref{algo:mapping} details the process of mapping the GPS location traces to OSM features.

For each datapoint, we add the OSM Feature ID, which we use to generate samples (\cref{subsec:Sample_Generation}) and later to calculate accurate travel distances (\cref{subsec:dataProcessing_Distance}).
We also add information from OpenStreetMap regarding the road, such as the type of the road, whether the road is one-way or two-way, whether it is a tunnel, etc. 
In our dataset, we encounter 14 different road types in total, which include  primary, residential, motorway, etc. 
For some roads, the type is ``\emph{unknown}' on {OpenStreetMap}, which we treat as a distinct type.

\SetKwInOut{Input}{Input}
% \SetKwInOut{Initialization}{INITIALIZATION}
\SetKwInOut{DATA}{Data}
\SetKwInOut{Output}{Output}
\SetKwInOut{Initialization}{Initialization}

\begin{algorithm}[h!]
\caption{Mapping Locations to OSM Features}\label{algo_filtering}
\label{algo:mapping}

\Input{$Locations \gets$ list of locations \\
$Map \gets$ OSM street-level map \\
$\textit{WINDOW} \gets$ lookahead and -back}
%\DATA{$Location$, $OSM\_Map$}
\Output{$Roads \rightarrow$ OSM features traveled}

\Initialization{

$NearbyRoads \gets [][]$ \tcc{list of nearby OSM features for each location}

$Roads \gets []$ \tcc{OSM feature for each location }

%$WINDOW$ $\leftarrow$ Previous and following 15 GPS locations

}

\For{$i \in \{1, \ldots, |Locations|\}$}{

$Nearby \gets Map.NearbyFeatures(Locations[i])$

$NearbyRoads[i] \gets Nearby$

}

\For{$i \in \{1, \ldots, |Locations|\}$}{

\If{$|NearbyRoads[i]| > 0$}{

$Frequency \leftarrow [\,]$

\For{$Road \in NearbyRoads[i])$}{
$Count \gets 0$

%\tcc{check within $WINDOW$}

\For{$j\! \in\! \{i\!-\!\textit{WINDOW}, \ldots, i\!+\!\textit{WINDOW}\}$}{

\For{$OtherRoad \in NearbyRoads[j]$}{

\If{$Road == OtherRoad$}{ $Count \gets Count+1$}
}

}

$Frequency[Road] \gets Count$

}

$Selected \gets argmax_{j}~Frequency[j]$

$Roads[i] \gets NearbyRoads[i][Selected]$

}

}

\end{algorithm} 

\subsection{Generating Samples}
\label{subsec:Sample_Generation}

Next, we generate a set of samples from the time series data by dividing the time series of each bus based on the traveled road segments.
Specifically, for each bus, we treat a maximal continuous travel on a particular road segment (i.e., particular OSM feature) as one sample. 
Each sample includes the starting datapoint, the ending datapoint, and the sum of the amount of energy or fuel used between them.

%We treat a maximal contiguous 
%After  mapping  the  locations of a bus to  roads,  we  segment the time series of each bus into disjoint contiguous samples based on road segments. 
%We identify a maximal continuous travel on a particular road segment (i.e., OSM feature) as one sample $s$.

\subsection{Calculating Travel Distance}
\label{subsec:dataProcessing_Distance}

Since GPS based locations are noisy, we combine them with OpenStreetMap to calculate the distance traveled for each sample accurately.
First, for each sample, we obtain the geometry of the corresponding road segment from OSM as a list of contiguous line segments.
%This list of geo-coordinates represent a list of line segments that 
%From OSM, we obtain the geometry of a road segment in the form of a list of geo-coordinates, that is, 
%
%    
%For calculating the distance of each sample, we first divide $G$ into multiple individual line segments, each having two endpoints.\footnote{A point consists of latitude and longitude.} 
Because the bus does not necessarily travel the complete distance of the road segment (e.g., it could turn on a different street before reaching the end of the road segment), we need to identify the first and last line segments that the bus actually traveled. 
We calculate the distance between each line segment and the starting and end points of the sample, which we denote $\textit{Dist}_S[]$ and $\textit{Dist}_E[]$, respectively.
Next, we identify the indices of the line segments that are closest to the starting and end points, which we denote $\textit{index}_S$ and $\textit{index}_E$, respectively.
Finally, we calculate the distance traveled for the sample based on the partial distance on line segment $\textit{index}_S$, the full distance of all line segments in between, and the partial distance on line segment $\textit{index}_E$, according to \cref{algo:distance}.
% on which the starting and endpoints of the same
% Then, we measure the distance between individual line segments and  $loc_0$ and $loc_n$ and store them in $Dist_1[\textrm{ }]$ and $Dist_2[\textrm{ }]$. Based on the minimum distance between line segments and points, we determine the index of the starting line $index_S$ and the starting line segment $line\textrm{ }0$. We also determine the index of the ending line $index_E$ and the ending line segment $line\textrm{ }n$. Then we calculate the distance of a sample using \cref{algo:distance}. 
% \Aron{``Algorithm'' (again, it might be easier to just use the cleveref package).}

\begin{algorithm}[t]
\caption{Calculating Travel Distance for Sample}
\label{algo:distance}
\DontPrintSemicolon
  
  \Input{$loc_S \gets$ starting point of sample\\
  $loc_E \gets$ end point of sample\\
   $line_1,\, line_2,\, \ldots,\, line_n \gets$ line segments of the OSM feature of the sample
  }
 
  %\DATA{}
   \Output{$L \rightarrow$ distance traveled}
   
   \For{$i \in \{1, \ldots, n\}$}{
   $\textit{Dist}_S[i] \gets$ distance($loc_S$, $line_i$) 
   
      $\textit{Dist}_E[i] \gets$ distance($loc_E$, $line_i$)
   }
   
   $\textit{index}_S \gets argmin_{i} ~ \textit{Dist}_S[i]$ %and $index_S \gets j$
   
  $\textit{index}_E \gets argmin_{i} ~ \textit{Dist}_E[i]$ %and $index_E \gets j$
   
  %\tcc{vehicle  moving in direction $line_{index_S}, line_{index_S+1}, \ldots, line_{index_E-1}, line_{index_E}$}
  \tcc{vehicle  moving in direction $line_{index_S},~ line_{index_S+1},~ \ldots,~ line_{index_E}$}
  
  \If{($index_S < index_E$)}
    {
         $l_1$ $\gets\!$ distance($loc_S$, second endpoint of $line_{index_S}$)
        
        $l_2$ $\gets\!$ sum length of $line_{index_S+1}, \ldots, line_{index_E-1}$
        
        $l_3$ $\gets\!$ distance(first endpoint of $line_{index_E}$, $loc_E$)
        
        $L \gets l_1 + l_2 + l_3$  
    }
    
    %\tcc{vehicle  moving in direction $line_{index_E}, line_{index_E+1}, \ldots, line_{index_S-1}, line_{index_S}$}
    \tcc{vehicle  moving in direction $line_{index_E},~ line_{index_E+1},~ \ldots,~  line_{index_S}$}
    
    \ElseIf{($index_S > index_E$)}
    {
    	$l_1$ $\gets\!$ distance($loc_E$, second endpoint of $line_{index_E}$)
           
        $l_2$ $\gets\!$ sum length of $line_{index_E+1}, \ldots, line_{index_S-1}$
           
        $l_3$ $\gets\!$ distance(first endpoint of $line_{index_S}$, $loc_S$)
        
        $L \gets l_1 + l_2 + l_3$
    }
    \tcc{$index_S == index_E$}
    \Else
    {
        $L \gets$ distance between $loc_S$ and $loc_E$.
    }
\end{algorithm}

\begin{comment}
\subsection{Energy/Fuel Usage Calculation}
For electric vehicles, we determine the energy consumption for each sample $s$ using the state of charge of the battery:
% For a sample $s \in S$, with $SoC_{startpoint}$ as the state of charge at the beginning of the trip and $SoC_{endpoint}$ as the state of charge at the end of the trip, 
$$\textit{Energy\,Consumption} = SoC_{\textit{endpoint}} - SoC_{\textit{startpoint}}$$
For fuel consumption calculation, we approach the same way using the total amount of fuel used over time:
% For a sample $s \in S$, with $fuelLevel_{startpoint}$ as the state of charge at the beginning of the trip and $fuelLevel_{endpoint}$ as the state of charge at the end of the trip, 
$$\textit{Fuel\,Consumption} = \textit{fuelLevel}_{\textit{endpoint}} - \textit{fuelLevel}_{\textit{startpoint}}$$
\end{comment}

\subsection{Removing Erroneous Samples}

Even though current and voltage values are almost always correctly recorded, we did find a few datapoints that have erroneous or missing values, which result in extremely low, negative energy consumption estimates. 
Note that many electric vehicles can recharge from braking; so energy consumption can in fact be negative for some shorter samples when the bus is slowing down or going downhill.
However, erroneous values result in implausibly low values.

%But we want to remove the samples where the energy consumption values are smaller than a certain number. 
%These outliers occurs when we have missing or erroneous voltage and/or current values. 

\begin{figure}[h!]
\pgfplotstableread[col sep=comma]{Data/Energy_Distribution_Electric.csv}\ElectricEnergy
\centering
    \begin{tikzpicture}[font=\scriptsize]
        \begin{axis}[
        SmallBarPlot,
            width=\linewidth,
            height = 4cm, 
            ymax = 55,
            xmin = -.25,xmax = 14.15,
            grid = major,
            xticklabels={{$\leq$-.2}, -.2,
            % -0.2\\to\\-0.15, 
            % -.15\\to\\-.1,
            % -.1\\to\\-.05,
            % -.05\\to\\-.025,
            % -.025\\to\\0,
            % 0\\to\\.025,
            % .025\\to\\.05,
            % .05\\to\\.1,
            % .1\\to\\.15,
            % .15\\to\\.2,
            % .2\\to\\.25,
            % .25\\to\\.3,
            % .3\\to\\.35,
            % .35\\to\\.4,
            -.15, -.1, -.05, -.025, 0, .025, .05, .1, .15, .2, .25, .3,
            {$\geq$.4}},
            yticklabel=\pgfmathprintnumber{\tick}\,$\%$,
            ylabel style = {align = center},
            xlabel= {Energy consumption [$\Delta SoC$]},
            ylabel = Fraction of samples
        ]
        \addplot [RedBars] table [x expr=\coordindex, y  = Energy_Electric] {\ElectricEnergy};
        \addplot[draw=red!60,thick,smooth] table [x expr=\coordindex, y  = Energy_Electric] {\ElectricEnergy};
        \addplot +[black!80, smooth, mark=none] coordinates {(1, 0) (1, 70)};
        \end{axis}
    \end{tikzpicture}
\caption{Distribution of energy consumption values for electric vehicle samples.}
\label{fig:Energy_Distribution}
\end{figure}
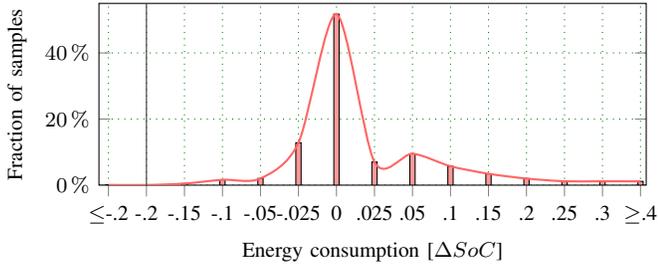

\cref{fig:Energy_Distribution} shows the distribution of the energy consumption values (measured as changes in $SoC$) for the 62,249 samples that we obtain for electric vehicles. 
Of these samples, 99.92\% have energy use values greater than or equal to -0.2. Only 50 samples have values lower than -0.2, constituting 0.08\% of the dataset. We remove these 50 samples from the~dataset.

\subsection{Incorporating Elevation} %, Weather, and Traffic Data}

To add road gradients to the samples, 
% we use the elevation map based on high-accuracy LiDAR data from the state government.\Aron{We should move the description of the datasource to the relevant section. There, we should specify exactly where we got the data.} 
we calculate the difference between the elevation at the start and end points of each sample.
The change in elevation captures the net potential energy gained or lost during the sample.
% and incorporate the dataset with the road gradients.

\subsection{Incorporating Weather}
Since our goal is to provide predictions for planning transit operations in advance, we cannot rely on real-time data for weather. 
Instead, we compute hourly weather predictions for each station based on the recorded historical weather data.
Then, for each sample, we compute the distance between the end point of the sample and each weather station, and we add the predicted weather features of the closest station to the sample.
%
% We use weather data collected from the Dark Sky API \cite{darkskyapi} at five-minute intervals. This data includes real-time temperature, humidity, visibility, pressure, wind speed, and precipitation etc. 
Our weather dataset has a number of features, of which we use
temperature (\emph{T}), humidity (\emph{H}), visibility (\emph{V}), wind speed (\emph{W}), and precipitation (\emph{P}).
%temperature, visibility, wind speed, humidity and precipitation.
%The weather dataset has data recorded in real-time for different stations in Chattanooga. It also has data recorded multiple times during an hour for a particular station. 
% For each sample~$s$, first, we determine the distance between the location of $s$ and different stations in Chattanooga. Then, we take the nearest station and calculate the hourly averages of various weather features for that station. Finally, we add the weather attributes information based on the time recorded for $s$. 

\subsection{Incorporating Traffic}

Our traffic dataset consists of timestamped speed values recorded for segments of roads in Chattanooga, which are identified using Traffic Message Channel (TMC) identifiers~\cite{tmc}. 
Each TMC segment represents a specific, directed segment of a major road, whose geometry is stored as a list of geo-points. 
While the TMC format is adequate for delivering and storing traffic information, we must also be able to integrate traffic data with our samples, which reference road segments using OSM Feature IDs.
%However, the traffic information has to be merged into a unifying schema that can be used with the other vehicle data sources in the models. 
To this end, we need to map OSM features to TMC segments.
%Therefore, the TMC segments are mapped to OpenStreetMap Ways \cite{osmways}.  
%
This mapping presents two challenges. % for two reasons.
First, OpenStreetMap typically divides roads into significantly smaller segments than TMC segments, so matching based on similarity of geometry is difficult.
Second, TMC segments cover only major roads, so most OSM features cannot be mapped to any TMC segment.
%This conversion therefore is a mapping of TMC road segments to OSM Ways, which also represent road segments. 
%In general, OSM Ways are significantly smaller than TMC segments. 

To set up the mapping, we first generate an OpenStreetMap routing graph.
This graph enables us to find the shortest driving-distance path between any two nodes, which represent real-world locations, returning a list of edges.
Each edge is labeled with the ID of the corresponding OSM feature (i.e., road segment).
%In this graph, the length is used as the edge weights, and each edge includes a labeled OSM Way associated with that edge. 
Next, for the start and end geo-points of each TMC segment, we find the closest nodes in the OSM routing graph.
%we find the nodes of the OSM  starting geo-point and ending geo-point are mapped to the closest OSM node, respectively. 
Finally, for each TMC segment, we find the shortest path in the OSM routing graph between the start and end nodes, and we map each edge (i.e., OSM feature) of the path to the TMC segment. % The shortest path between the start node and end node is found, and each OSM edge in the path is labeled with the TMC id accordingly. 

However, in some cases, the start and end geo-points of a TMC segment are matched to OSM nodes on the opposite sides of a road, which causes errors in the mapping. 
Therefore, instead of finding only the nearest OSM node, we find the four nearest nodes for each start and end geo-point. 
Then, we find all the shortest paths between all the start and end nodes, select the path whose length matches the actual length of the TMC segment most closely, and map the OSM features of only this path to the TMC segment. %the path whose length matches  For a given start node, routes are found to each of the possible end node candidates. The length of each route is compared to the actual length of the road segment, and the route with the length closest to the actual TMC segment length is taken as the true route. 
We found that this process significantly improves the OSM to TMC mapping.

Based on this mapping, we add traffic information to our samples. 
Similar to weather, we cannot rely on real-time traffic for energy use prediction.
Instead, we compute average traffic conditions for each TMC segment for each hour of each day of the week based on the recorded data, and we use these hourly averages as traffic predictions.
For each sample, we add the hourly prediction for the TMC segment to which the OSM feature of the sample is mapped.
For samples that cannot be mapped, we impute special values, which we discuss below.
We add two features from our traffic dataset to each sample: \emph{speed ratio} and \emph{jam factor}. Speed ratio is the actual traffic speed over the free-flow speed; values around 1 mean light or no traffic, while values around 0 mean very heavy traffic.
Jam factor indicates the expected quality of travel, ranging from 0 (light or no traffic) to 10 (road closure) \cite{hereapi}.
For samples that cannot be mapped to a TMC segment, we let the speed ratio and jam factor be 1 and 0, respectively,
since road segments that are missing from our traffic dataset are typically minor roads, which rarely experience heavy traffic.

\section{Energy Consumption Prediction Models}
\label{sec:predictionModels}

We apply three different machine-learning models for predicting energy consumption: artificial neural network, linear regression, and decision tree regression.
% it is better to explain why choose these when we introduce them the first time
We chose neural networks for their superior prediction performance, which is confirmed by our numerical results.
In contrast, linear and decision tree regression do not perform as well, but their results are easier to understand and explain.
For example, linear regression shows the direct relation between input variables and target features.

We map categorical variables (e.g., road type) into sets of binary features using one-hot encoding.
We train all three models to minimize \emph{mean squared error} (MSE).

\subsection{Artificial Neural Network}
We found that different network structures work best for diesel and electric vehicles. 
%The artificial neural network models for both diesel and electric have different structures. 
For electric vehicles, the best model has one input, two hidden, and one output layer. The input layer has one neuron for each predictor variable. The two hidden layers have 100 neurons and 80 neurons, respectively. For diesel, the best model has one input, five hidden, and one output layer. The five hidden layers have 400, 200, 100, 50, and 25 neurons, respectively. In all the hidden layers, we use sigmoid activation, and we use linear activation in the output layer. We optimize the models using the \emph{Adam} optimizer~\cite{kingma2014adam} with learning rate 0.001.

\subsection{Linear Regression}
Our second model is a standard multiple linear regression.
%Linear regression fits a linear model with coefficients to minimize the residual sum of squares between the observed targets and the targets predicted by the linear approximation.
%We evaluate the Linear Regression model based on $R^2$ value and \emph{Mean Squared Error (MSE)}.

\subsection{Decision Tree}
\Aron{what library did you use?}
Our third model is % supervised learning technique that we use for the regression problem is 
decision tree regression~\cite{dtr}. 
This model builds a tree structure based on the training samples, where each node represents a decision based on the value of a feature variable, and leaf nodes provide predictions.
We use the implementation provided by the \emph{scikit-learn} Python library.
% %This model observes the characteristics of an object and trains a model in a tree-like structure to predict meaningful outputs. 

\section{Numerical Results}
\label{sec:numerical}

%We describe the results of different algorithms in this section.
%After preparing the samples for both the electric and diesel buses, we split them into randomized training ($80\%$) and test sets ($20\%$).
%For both electric and diesel buses, we have a set of 26 features, including energy use as the target feature. 
%
%To enhance the accuracy of prediction, we omit some features based on our numerical results (\cref{sec:featureComp1,sec:featureComp2}). %predictor variables are dropped/added based on practical experience; adding/dropping of some variables increased efficiency. 
%Finally, 

\subsection{Mapping GPS Locations to Road Segments}

\begin{figure}[t]
\minipage{\linewidth}
\centering
  \includegraphics[width=0.48\linewidth,height = 2.4cm]{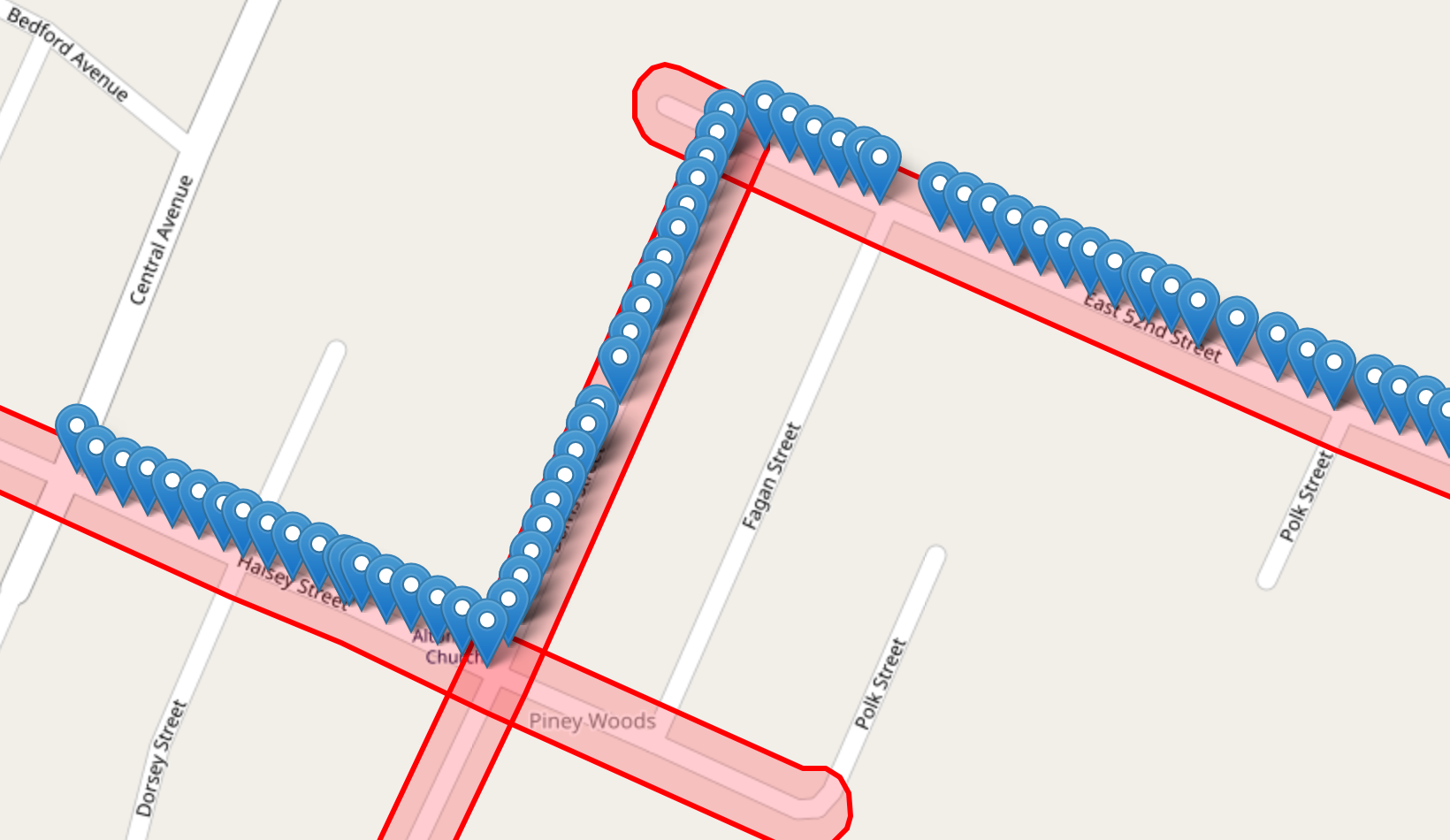}
  \caption{Mapping without noise.}\label{fig:Ground_Truth}
\endminipage

\minipage{0.48\linewidth}
  \includegraphics[width=\linewidth,height = 2.4cm]{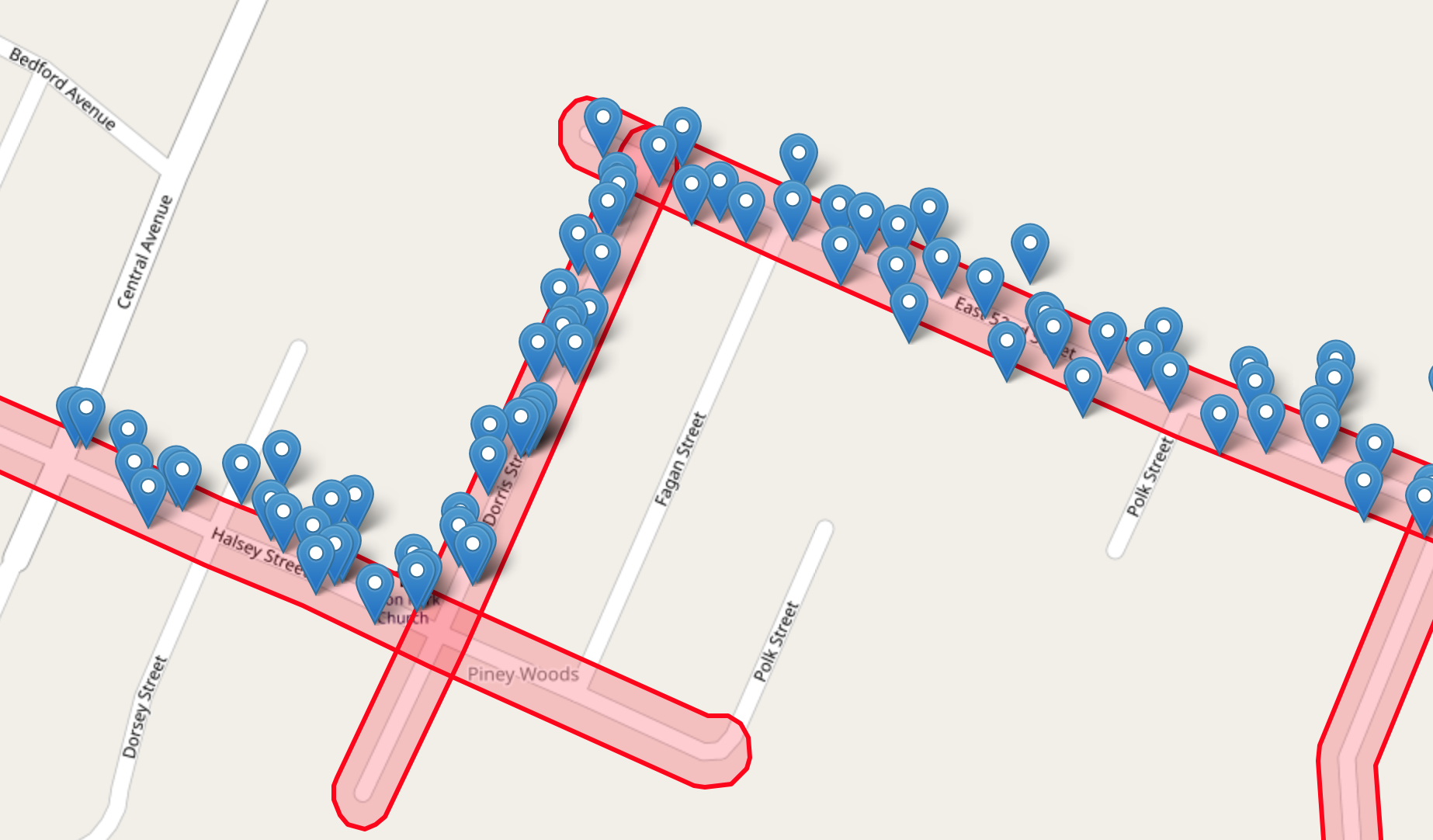}
  \caption{Mapping with noise with 14-meter std.\ dev.}\label{fig:noise0_000125}
\endminipage\hfill
\minipage{0.48\linewidth}%
  \includegraphics[width=\linewidth,height = 2.4cm]{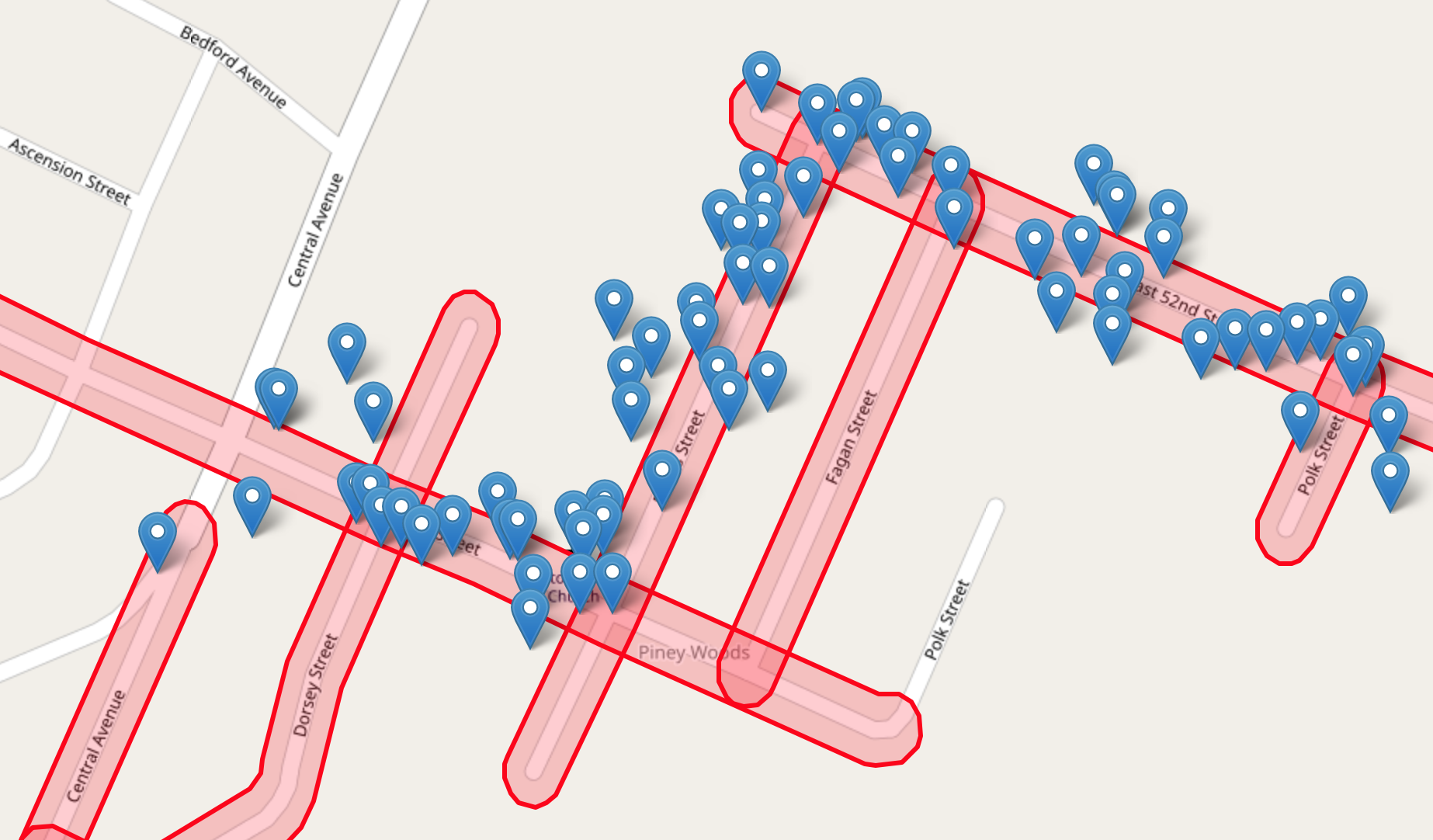}
  \caption{Mapping with noise with 28-meter std.\ dev.}\label{fig:noise0_00025}
\endminipage
\end{figure}

We begin by evaluating the accuracy of our algorithm for mapping noisy locations to OSM features (\cref{algo:mapping}).
Since we do not have ground truth for the correct mapping in our GPS-based dataset, we create a test dataset with known ground truth.
First, we generate routes using a street-level map and select a set of locations along these routes, which are precisely on the roads (\cref{fig:Ground_Truth}).
%To evaluate the performance of our algorithm discussed in \cref{subsec:dataProcessing_mapping_GPS}, we use coordinates on specific routes based on street-level maps with  locations placed precisely on the roads.
%We plot these coordinates to real streets using the algorithm and consider that as our ground truth. 
Then, we add random noise to these locations, generated using a two-dimensional Gaussian distribution with zero mean. 
We vary the 
% from \Aron{I think that it will be easier for the reader if we give values in meters immediately.}$0.00001$ degrees to $0.001$ degrees. For the city of Chattanooga, $1$ degree in the latitude and longitude represent approximately \Aron{Then, we can skip this part.} $111$ kilometers and $91$ kilometers. So we multiplied the scale of noise by $1.2$ for latitude and $1$ for longitude. So, the noisy coordinates are located 
standard deviation of the noise between $1$ meter and $110$ meters in both directions (\cref{fig:noise0_000125,fig:noise0_00025}). 
Finally, we map the noisy locations to road segments using \cref{algo:mapping} and measure accuracy as the ratio of correctly mapped locations. % of correct mapping.  and then determine the accuracy of mapping locations to correct streets. 

% \begin{figure}[!htb]
% \minipage{0.45\textwidth}
% \centering
%   \includegraphics[width=\linewidth,height = 4cm]{figures/GroundTruth.png}
%   \caption{Mapping without noise}\label{fig:Ground_Truth}
% \endminipage

% \minipage{0.25\textwidth}
%   \includegraphics[width=\linewidth,height = 3cm]{figures/noise0_000125.png}
%   \caption{Mapping with noise in 14 meter radius}\label{fig:noise0_000125}
% \endminipage\hfill
% \minipage{0.23\textwidth}%
%   \includegraphics[width=\linewidth,height = 3cm]{figures/noise0_00025.png}
%   \caption{Mapping with noise in 28 meter radius}\label{fig:noise0_00025}
% \endminipage
% \end{figure}

\cref{fig:Ground_Truth,fig:noise0_000125,fig:noise0_00025} show locations with different levels of noise added, highlighting in red the road segments to which locations are mapped by \cref{algo:mapping}.
\cref{fig:algorithm_results} shows the accuracy of mapping with various levels of noise, ranging from zero to 110-meter standard deviation in both directions.
As expected, the accuracy of the algorithm decreases as the level of noise increases.
However, for reasonable noise levels, it performs very well: 
with 14-meter standard deviation, it can still correctly map 84.5\% of locations.

% The algorithm performs very well
% , we plot the coordinates before introducing any noise. In \cref{}, we can see that the coordinates are scattered around the place due to the introduction of noise with $14$-meter and $28$-meter standard deviation, respectively. Still, the algorithm maps the coordinates with noise with $14$ meters standard deviation correctly with an accuracy of $84.5\%$. But in \cref{fig:noise0_00025}, the accuracy drops to $48.5\%$.

\pgfplotstableread[col sep=comma]{Data/Noise_Results.csv}\noise

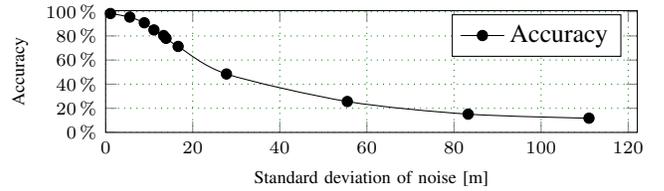
\begin{figure}[t]
    \centering
    \resizebox{\columnwidth}{!}{
    \begin{tikzpicture}
        \begin{axis}
        [
            ylabel=Accuracy,
            xlabel={Standard deviation of noise [m]},
            ylabel style = {align = center},
            	grid=major,
            	tick label style={font=\scriptsize}, 
            % xtick = data,
            % xticklabels from table={\noise}{Noise},
            % yticklabels from table={\noise}{Accuracy},
            yticklabel=\pgfmathprintnumber{\tick}\,$\%$,
            xticklabel=\pgfmathprintnumber{\tick},
            legend pos=north east,
            label style = {align = center,font=\scriptsize},
            width=\linewidth,
            height=3.25cm,
            xmin = 0,
            ymin = 0, ymax = 101
        ]
            \addplot [black, smooth, mark=*] table [x=Distance_in_Meter, y=Accuracy] {\noise};
            \addlegendentry{Accuracy}
        \end{axis}
    \end{tikzpicture}
    }
    \caption{Accuracy of mapping noisy locations to road segments.}
    \label{fig:algorithm_results}
\end{figure}

% \cref{fig:algorithm_results} shows that the algorithm degrades in performance with varying amounts of noise. 

\subsection{Comparison of Weather, Traffic and Elevation Features}
\label{sec:featureComp1}

For both electric and diesel buses, we have a set of 26 features in each sample, besides energy use as the target feature. 
Now, we study which of these features are the most useful for predicting energy use, and which subset of features results in the lowest prediction error.

%To enhance the accuracy of prediction, we omit some features.

After preparing the samples for both electric and diesel buses, we randomly split them into training (80\%) and test sets (20\%). 
We use the same split ratio in all subsequent experiments.
Since  neural networks attain the lowest prediction error (see \cref{sec:compPredMod}), we compare features based on this model. 
We include vehicle-level data in all experiments, and try different combinations of weather, elevation, and traffic~data.

\pgfplotstableread[col sep=comma]{Data/Different_Features_Electric.csv}\FeatureElectric
\pgfplotstableread[col sep=comma]{Data/Different_Features_Diesel.csv}\FeatureDiesel
\begin{figure}[t]
\begin{subfigure}[b]{\linewidth}
    \centering
    \begin{tikzpicture}[font=\scriptsize]
    \begin{axis}[
            width = \linewidth,
            height = 2.8cm, 
            xlabel=Features used for prediction,
            xtick=data,
            ymin = 0.0310,
            xmin = -.15,
            xmax = 4.1,
            grid = major,
            xticklabels=
            {{\emph{Weather}}, {\emph{Traffic}}, {\emph{Elevation}}, {\emph{Traffic+Elevation}}, {\emph{All}}},
            label style = {align = center},
            % xlabel=Weather Features,
            ylabel = Error in MSE
        ]
        \addplot [only marks, red!60, smooth, mark=*] table [x expr=\coordindex, y={Loss}] {\FeatureElectric};
        % \addlegendentry{Electric}
\end{axis}
\end{tikzpicture}
    \caption{Electric}
    \label{fig:Electric_DifferentFeatures}
\end{subfigure}
\begin{subfigure}[b]{\linewidth}
\begin{tikzpicture}[font=\scriptsize]
\begin{axis}[
            width = \linewidth,
            height = 2.8cm, 
            xlabel=Features used for prediction,
            xtick=data,
            ymin = 0.00069,
            xmin = -.15,
            xmax = 4.1,
            grid = major,
            % xticklabels=
            % {T, H, {$V$}, {$P$}, {$W$}, All, None,{$TVPH$},{$TVP$}},
            xticklabels=
            {{\emph{Weather}}, {\emph{Traffic}},{\emph{Elevation}}, {\emph{Traffic+Elevation}}, {\emph{All}}},
            % xtick label style={font=\footnotesize},  
            label style = {align = center},
            % xlabel=Weather Features,
            ylabel = Error in MSE
        ]
        \addplot [only marks, blue!60, smooth, mark=*] table [x expr=\coordindex, y={Loss}] {\FeatureDiesel};
        % \addlegendentry{Diesel}
\end{axis}
\end{tikzpicture}
    \caption{Diesel}
    \label{fig:Diesel_DifferentFeatures}
\end{subfigure}
\caption{Prediction error with various sets of features. Note that electric and diesel energy are measured in different units.}
\label{fig:DifferentFeatures_Comparison}
\end{figure}
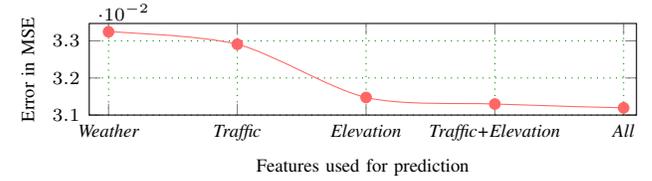
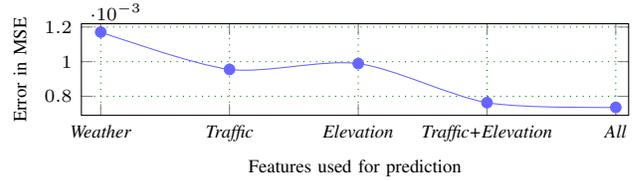

%To identify the optimal set of features, we provide a thorough comparison among various factors and different combinations of those factors. 
%for the prediction model. 
\Aron{this should be a bar plot}
\cref{fig:Electric_DifferentFeatures} shows that elevation is by far the most significant feature for electric vehicles.
Traffic data does improve prediction, but its impact is much smaller, especially if elevation is already included. This can be explained by regenerative breaking: the energy use of electric vehicles is not impacted by heavy traffic since they do not lose energy due to frequent braking. 
%does not help much with energy use prediction, which makes sense because practically electric vehicles perform better in heavy traffic due to regenerative braking. 
%
\Aron{this should be a bar plot}
On the other hand, \cref{fig:Diesel_DifferentFeatures} shows that for diesel vehicles, both elevation and traffic data are significant, and both need to be included for good performance.
Finally, we find that weather data has the lowest impact on prediction error for both electric and diesel vehicles.

%, when we incorporate our dataset with traffic data, the prediction result seems to improve. Adding elevation also helps with the prediction performance for both buses. According to the \emph{MSE} produced by each of the features, we see that all three features together provide the least \emph{MSE} score for both electric and diesel buses.

\subsection{Comparison of Different Weather Features}
\label{sec:featureComp2}

Since weather data has many features, we also present a comparison among various weather features to see which ones help with prediction the most. We consider temperature (\emph{T}), humidity (\emph{H}), visibility (\emph{V}), wind speed (\emph{W}), and precipitation (\emph{P}) in this comparison.

%We also check the different combinations of these features to identify which combination helps to produce a better result.

\pgfplotstableread[col sep=comma]{Data/Weather_Features_Electric.csv}\weatherFeatureElectric
\pgfplotstableread[col sep=comma]{Data/Weather_Features_Diesel.csv}\weatherFeatureDiesel
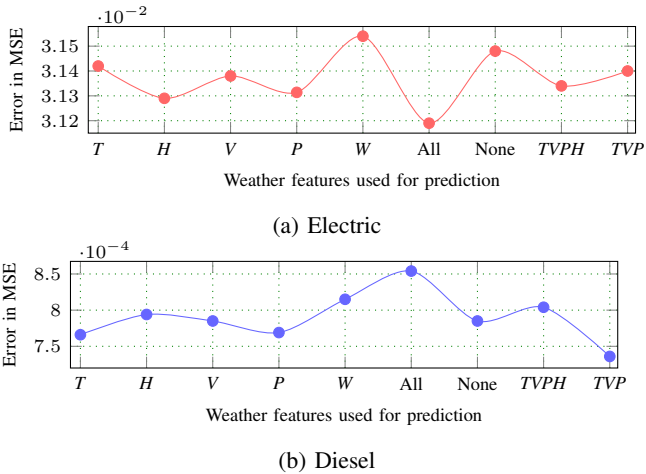
\begin{figure}[t]
\begin{subfigure}[b]{\linewidth}
    \centering
    \begin{tikzpicture}[font=\scriptsize]
    \begin{axis}[
            width = \linewidth,
            height = 3.0cm, 
            xtick=data,
            xlabel=Weather features used for prediction,
            ymin = 0.03115,
            xmin = -.15,
            xmax = 8.12,
            grid = major,
            xticklabels=
            {{\emph{T}}, {\emph{H}}, {\emph{V}}, {\emph{P}}, {\emph{W}}, All, None,{\emph{TVPH}},{\emph{TVP}}},
            label style = {align = center},
            legend style={at={(0,1)},anchor=north west,font=\tiny},
            % xlabel=Weather Features,
            ylabel = Error in MSE
        ]
        \addplot [red!60, smooth, mark=*] table [x expr=\coordindex, y={Loss}] {\weatherFeatureElectric};
        %\addlegendentry{Electric}
\end{axis}
\end{tikzpicture}
    \caption{Electric}
    \label{fig:Electric_WeatherFeatures}
\end{subfigure}
\begin{subfigure}[b]{\linewidth}
\begin{tikzpicture}[font=\scriptsize]
\begin{axis}[
            width = \linewidth,
            height = 3.0cm, 
            xtick=data,
            xlabel=Weather features used for prediction,
            ymin = 0.000720,
            xmin = -.15,
            xmax = 8.12,
            grid = major,
            % xticklabels=
            % {T, H, {$V$}, {$P$}, {$W$}, All, None,{$TVPH$},{$TVP$}},
            xticklabels=
            {{\emph{T}}, {\emph{H}}, {\emph{V}}, {\emph{P}}, {\emph{W}}, All, None,{\emph{TVPH}},{\emph{TVP}}},
            % xtick label style={font=\footnotesize},  
            label style = {align = center},
            legend style={at={(0,1)},anchor=north west,font=\tiny},
            % xlabel=Weather Features,
            ylabel = Error in MSE
        ]
        \addplot [blue!60, smooth, mark=*] table [x expr=\coordindex, y={Loss}] {\weatherFeatureDiesel};
        %\addlegendentry{Diesel}
\end{axis}
\end{tikzpicture}
    \caption{Diesel}
    \label{fig:Diesel_WeatherFeatures}
\end{subfigure}
\caption{Prediction error with various weather features.}
\label{fig:Weather_Features_Comparison}
\end{figure}

\cref{fig:Weather_Features_Comparison} shows prediction error with various combinations of weather features (with traffic and elevation always included).
For electric vehicles, we attain lowest error when we use all five features together (\cref{fig:Electric_WeatherFeatures}). 
On the other hand, for diesel vehicles, we attain lowest error using only three features: temperature, visibility and precipitation (\cref{fig:Diesel_WeatherFeatures}).
This may be explained by over-fitting when using more features.

\subsection{Comparison of Prediction Models for Samples}
\label{sec:compPredMod}

We first evaluate the three machine-learning models based on how well they predict energy use for samples. 
Our samples represent segments of trips that are short in both distance and duration, presenting a challenging problem for prediction. 

\cref{fig:Electric_Model_Comparison,fig:Diesel_Model_Comparison} show \emph{mean squared error} (MSE) and \emph{mean absolute error} (MAE) for the three models.
Based on MSE, the artificial neural network (ANN) outperforms the other two models for both electric and diesel vehicles.
However, based on MAE, ANN outperforms decision trees (DT) for diesel vehicles but not for electric vehicles.
Note that we optimized all models to minimize MSE, which can explain the slightly inferior performance of ANN for MAE. 
We have not encountered any overfitting since our training and testing errors were consistent for each model.
% the artificial neural network has better performance accuracy than the other two regression models.
% For electric vehicles, even though the ANN has the least \emph{MSE} but the decision tree has the least mean absolute error (see ).
% But for diesel vehicles, the artificial neural network has the least \emph{MSE} and \emph{MAE} among all the regressor models (see \cref{fig:Diesel_Model_Comparison}).
% We pick \emph{MSE} over \emph{MAE} for optimizing our models beacuse \emph{MSE} is more sensitive to outliars than \emph{MAE}.

\pgfplotstableread[col sep=comma]{Data/Diesel_Vehicles_Shorter_Interval_Loss.csv}\Diesel
\pgfplotstableread[col sep=comma]{Data/Electric_Vehicles_Shorter_Interval_Loss.csv}\Electric
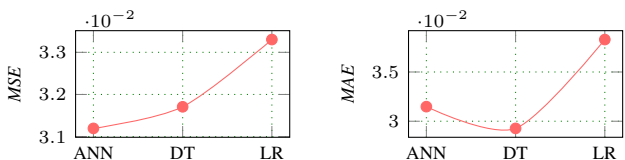
\begin{figure}[t]
\begin{subfigure}[b]{0.5\linewidth}
    \centering
    \begin{tikzpicture}[font=\scriptsize]
    \begin{axis}
         [
        % SmallBarPlot,
            % xlabel=Different regression models,
	       ylabel=\emph{MSE},
            ylabel style = {align = center},
            	grid=major,
            xtick = data,
            xticklabels from table={\Electric}{Models},
            yticklabel=\pgfmathprintnumber{\tick},
            tick label style={font=\scriptsize}, 
            legend pos=north east,
            width=\linewidth,
            height=3cm,
            legend style={at={(1,1)},anchor=north east,font=\scriptsize}
        ]
            \addplot [red!60, smooth, mark=*] table [x expr=\coordindex, y=MSE] {\Electric};
            % \addplot [BlueBars] table [x expr=\coordindex, y  = MAE] {\Electric};
            % \addlegendentry{Deep Neural Network}
            % \addplot [black, smooth, mark=square*] table [x expr=\coordindex, y=MAE] {\Electric};
            % \addlegendentry{Decision Tree}
            % \addplot [green!60, smooth, mark=triangle] table [x expr=\coordindex, y=LR] {\Electric};
            % \addlegendentry{Linear Regression}
    \end{axis}
\end{tikzpicture}
\label{Electric_MSE}
    % \caption{Electric}
    % \label{fig:Electric_ShorterTrips}
\end{subfigure}%
\begin{subfigure}[b]{0.5\linewidth}
    \centering
    \begin{tikzpicture}[font=\scriptsize]
    \begin{axis}
         [
        % SmallBarPlot,
            % xlabel=Different regression models,
	       ylabel=\emph{MAE},
            ylabel style = {align = center},
            	grid=major,
            xtick = data,
            xticklabels from table={\Electric}{Models},
            yticklabel=\pgfmathprintnumber{\tick},
            tick label style={font=\scriptsize}, 
            legend pos=north east,
            width=\linewidth,
            height=3cm,
            legend style={at={(1,1)},anchor=north east,font=\scriptsize}
        ]
            \addplot [red!60, smooth, mark=*] table [x expr=\coordindex, y=MAE] {\Electric};
            % \addplot [BlueBars] table [x expr=\coordindex, y  = MAE] {\Electric};
            % \addlegendentry{Deep Neural Network}
            % \addplot [black, smooth, mark=square*] table [x expr=\coordindex, y=MAE] {\Electric};
            % \addlegendentry{Decision Tree}
            % \addplot [green!60, smooth, mark=triangle] table [x expr=\coordindex, y=LR] {\Electric};
            % \addlegendentry{Linear Regression}
    \end{axis}
\end{tikzpicture}
\label{Electric_MAE}
\end{subfigure}%

\caption{\emph{Mean square error} (MSE) and \emph{mean absolute error} (MAE) for electric vehicle samples.}
\label{fig:Electric_Model_Comparison}
\end{figure}

%%%%%%%%%%%%%%%%%%%%%%%%%%%%%%%%%%%%%%%%%%%%%%%%%%%%%%%%%%
%%%%%%%%%%%%%%%%%%%%%%%%%%%%%%%%%%%%%%%%%%%%%%%%%%%%%%%%%%
%%%%%%%%%%%%%%%%%%%%%%%%%%%%%%%%%%%%%%%%%%%%%%%%%%%%%%%%%%
\begin{figure}[t]
\begin{subfigure}[b]{0.5\linewidth}
\begin{tikzpicture}[font=\scriptsize]
    \begin{axis}
        [
            % xlabel=Different regression models,
	       ylabel=\emph{MSE},
            ylabel style = {align = center},
            	grid=major,
            xtick = data,
            xticklabels from table={\Diesel}{Models},
            yticklabel=\pgfmathprintnumber{\tick},
            tick label style={font=\scriptsize}, 
            legend pos=north east,
            width=\linewidth,
            height=3cm,
            legend style={at={(1,1)},anchor=north east,font=\scriptsize}
        ]
            \addplot [Blue!60, smooth, mark=*] table [x expr=\coordindex, y=MSE] {\Diesel};
            % \addlegendentry{Deep Neural Network}
            % \addplot [black, smooth, mark=square*] table [x expr=\coordindex, y=MAE] {\Diesel};
        \end{axis}
\end{tikzpicture}

    % \caption{Diesel}
    \label{fig:Diesel_MSE}
\end{subfigure}%
\begin{subfigure}[b]{0.5\linewidth}
\begin{tikzpicture}[font=\scriptsize]
    \begin{axis}
        [
            % xlabel=Different regression models,
	       ylabel=\emph{MAE},
            ylabel style = {align = center},
            	grid=major,
            xtick = data,
            xticklabels from table={\Diesel}{Models},
            yticklabel=\pgfmathprintnumber{\tick},
            tick label style={font=\scriptsize}, 
            legend pos=north east,
            width=\linewidth,
            height=3cm,
            legend style={at={(1,1)},anchor=north east,font=\scriptsize}
        ]
            \addplot [Blue!60, smooth, mark=*] table [x expr=\coordindex, y=MAE] {\Diesel};
            % \addlegendentry{Deep Neural Network}
            % \addplot [black, smooth, mark=square*] table [x expr=\coordindex, y=MAE] {\Diesel};
        \end{axis}
\end{tikzpicture}

    % \caption{Diesel}
    \label{fig:Diesel_MAE}
\end{subfigure}
\caption{\emph{Mean square error} (MSE) and \emph{mean absolute error} (MAE) for diesel vehicle samples.}
\label{fig:Diesel_Model_Comparison}
\end{figure}
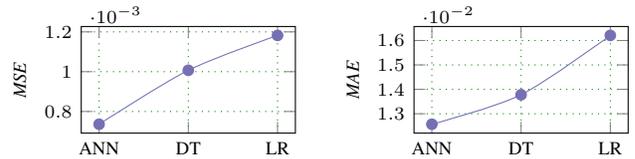

\subsection{Comparison of Prediction Models for Longer Trips}

Finally, we study how well our models perform with respect to predicting energy use for longer trips.
First, we divide our time series into longer trips, varying the length of the trips between 10 minutes and 6 hours.
For each trip, we generate a set of samples (as described in \cref{sec:dataProcessing}), use our models to predict energy use for each sample, and then compare the sum of these predictions to the actual energy use of the trip.

\cref{fig:LongerTrips_Prediction} shows the relative prediction error for trips of various lengths. 
For each length, we plot an average error value computed over many trips.
We see that relative prediction error is generally lower for longer trips; this is expected as the individual errors of large numbers of samples cancel each other out with an unbiased prediction model.
For diesel vehicles, we find that the ANN outperforms the other models significantly for all trip lengths.
On the other hand, for electric vehicles, ANN and DT perform equally well for most trip lengths.

% We see that prediction error decreases as the length of the 

% The training samples are consist of very short duration trips taken by the buses on a particular road segment. To identify how the regressor models work for longer trips, we divide our test set into tours with longer durations ranging from 10 minutes to 6 hours. 
% We plot the errors in prediction for longer trips for all the models in . In both cases, the error lessens as the trip duration increases. For diesel buses in \cref{fig:Diesel_LongerTrips}, still, the performance of ANN is better than the other regressors. But for electric buses in \cref{fig:Electric_LongerTrips}, ANN and Decision Tree has a similar reduction in errors.

 \pgfplotstableread[col sep=comma]{Data/Diesel_Vehicles_Longer_Interval_Loss.csv}\Diesel
\pgfplotstableread[col sep=comma]{Data/Electric_Vehicles_Longer_Interval_Loss.csv}\Electric
\begin{figure}[t]
\begin{subfigure}[b]{\linewidth}
    \centering
    \begin{tikzpicture}[font=\scriptsize]
    \begin{axis}
        [
            xlabel=Trip duration in minutes,
	       ylabel=Error,
            ylabel style = {align = center},
            	grid=major,
            xtick = data,
            xticklabels=
            {10,20,30,40,50,60,120,180,240,300,360},
            yticklabel=\pgfmathprintnumber{\tick}\,$\%$,
            tick label style={font=\scriptsize}, 
            legend pos=north east,
            width=\linewidth,
            height=3.2cm,
            xmin = -0.5,xmax = 11,
            ymin = 9, ymax = 40,
            legend style={at={(1,1)},anchor=north east,font=\scriptsize}
        ]
            \addplot [red!80, smooth, mark=*] table [x expr=\coordindex, y=FF] {\Electric};
            \addlegendentry{ANN}
            \addplot [Black, smooth, mark=square*] table [x expr=\coordindex, y=DT] {\Electric};
            \addlegendentry{DT}
            \addplot [Green!80, smooth, mark=triangle*] table [x expr=\coordindex, y=LR] {\Electric};
            \addlegendentry{LR}
    \end{axis}
\end{tikzpicture}
    \caption{Electric}
    \label{fig:Electric_LongerTrips}
\end{subfigure}
\begin{subfigure}[b]{\linewidth}
\begin{tikzpicture}[font=\scriptsize]
    \begin{axis}
        [
            xlabel=Trip duration in minutes,
	       ylabel=Error,
            ylabel style = {align = center},
            	grid=major,
            xtick = data,
            xticklabels=
            {10,20,30,40,50,60,120,180,240,300,360},
            yticklabel=\pgfmathprintnumber{\tick}\,$\%$,
            tick label style={font=\scriptsize}, 
            legend pos=north east,
            width=\linewidth,
            height=3.2cm,
            xmin = -0.5,xmax = 11,
            ymin = 1, ymax = 20,
            legend style={at={(1,1)},anchor=north east,font=\scriptsize}
        ]
            \addplot [red!80, smooth, mark=*] table [x expr=\coordindex, y=FF] {\Diesel};
            \addlegendentry{ANN}
            \addplot [Black, smooth, mark=square*] table [x expr=\coordindex, y=DT] {\Diesel};
            \addlegendentry{DT}
            \addplot [Green!80, smooth, mark=triangle*] table [x expr=\coordindex, y=LR] {\Diesel};
            \addlegendentry{LR}
        \end{axis}
\end{tikzpicture}

    \caption{Diesel}
    \label{fig:Diesel_LongerTrips}
\end{subfigure}
\caption{Prediction error for longer trips with neural network (ANN), decision tree (DT), and linear regression~(LR).}
\label{fig:LongerTrips_Prediction}
\vspace{-0.1in}
\end{figure}
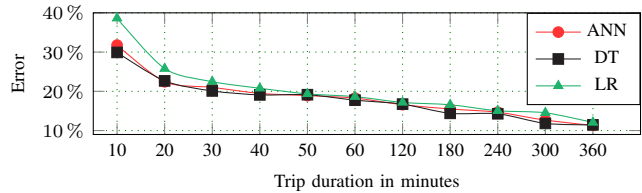
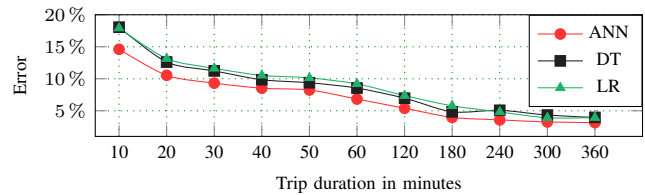

\section{Related Work}
\label{sec:related}

Our study is most closely related to the work of Cauwer et al. and of Wickramanayake and Bandara.
Cauwer et al. \cite{de2017data} use a cascade of ANN and multiple linear regression models as a data-driven energy-consumption prediction method for EVs.  
Their study uses vehicle monitoring data as time series of tuples for two types of vehicles with location, vehicle speed, and energy-consumption information, such as battery voltage, current, and SoC. Their dataset also includes road network data, weather data, and an altitude map. Our approach has some similarity to this study. However, we also use traffic data in our model, which we find to be very helpful with diesel prediction.
Wickramanayake and Bandara \cite{wickramanayake2016fuel} assess three different techniques for fuel consumption prediction of a long-distance public bus. Their time series tuples include GPS location, bearing, elevation, distance travelled, speed, acceleration, ignition status, battery voltage, fuel level, and fuel consumption. The authors compare the performance among two ensemble models, random forest and gradient boosting, and one ANN model. However, their study lacks critical parameters, such as road information, traffic, weather, etc. 

Perrotta et al. \cite{perrotta2017application} compare the performance of SVM, RF, and ANN in modelling fuel consumption of a large fleet of trucks. Their features include gross vehicle weight, speed, acceleration, geographical position, torque percentage, revolutions of the engine, activation of cruise control, use of brakes and acceleration pedal, measurement of travelled distance, fuel consumption. The study also combines some road characteristics%
%, such as road gradient, the radius of curvature of the road, measurements of the road unevenness, and the macrotexture of the road surface
. From the comparison of the RMSE, MAE, and R2 score of the prediction, RF gives the best performance. 
Nageshrao et al. \cite{nageshrao2017charging} models the energy consumption of electric buses based on time-dependent factors such as ambient temperature and speed, battery capacity, total mass, battery parameters, etc. They use a NARX based ANN time series predictor to predict the state of charge of the battery.
Gao et al. \cite{gao2018electric} discuss an adaptive wavelet neural network (WNN) based energy prediction. The study uses features such as day type, temperature, rainfall, the travelled distance, and clarity of the atmosphere. The study groups the trip days based on similar attributes, using Grey Relational Analysis (GRA) and then implements the Adaptive WNN.

Some researchers propose methods for optimizing the operations of vehicle fleets.
Wang et al. \cite{wang2018bcharge} design a real-time charging scheduling system, called bCharge, for electric bus fleets. They implement the system with the real-world streaming dataset from Shenzhen, China, including GPS data, bus stop data, bus transaction lines, bus charging station data, and electricity rate~data.
Murphey et al. \cite{murphey2012intelligent} propose the ML\_EMO\_HEV framework for energy management optimization in an hybrid-electric vehicles. Their framework first uses a ANN to model the road environment of a driving trip as a sequence of different roadway types and traffic congestion levels. Then, it uses an additional ANN to model the driver's instantaneous reaction to the driving environment. Finally, the framework uses an additional set of ANN to emulate the optimal energy management strategy.

\section{Discussion and Conclusion}
\label{sec:concl}

We presented a framework for the data-driven prediction of the energy use of electric and internal-combustion vehicles, which we evaluated on real-world data collected from a transit fleet.
Our results show that it is possible to collect, aggregate, and process heterogeneous transit data effectively.
We found that generally, artificial neural networks perform best for predicting energy use.
For diesel buses, we achieve best results using 21 predictor variables: % to forecast energy consumption.
%The variables are 
travel distance, 14 road-type features, elevation change, 3 weather features, and 2 traffic features.
For electric buses, we achieve best results using 23 predictor variables, which include 2 more weather features.
We also found that relative prediction error is lower for longer trips, which facilitates the long-term planning of transit operations.

\paragraph*{Acknowledgment}
We thank the anonymous reviewers.
This material is based upon work supported by the Department of Energy, Office of Energy Efficiency and Renewable Energy (EERE), under Award Number DE-EE\,0008467.
%Disclaimer: This report was prepared as an account of work sponsored by an agency of the United States Government. Neither the United States Government nor any agency thereof, nor any of their employees, makes any warranty, express or implied, or assumes any legal liability or responsibility for the accuracy, completeness, or usefulness of any information, apparatus, product, or process disclosed, or represents that its use would not infringe privately owned rights. 
%Reference herein to any specific commercial product, process, or service by trade name, trademark, manufacturer, or otherwise does not necessarily constitute or imply its endorsement, recommendation, or favoring by the United States Government or any agency thereof. 
%The views and opinions of authors expressed herein do not necessarily state or reflect those of the United States Government or any agency thereof.

\bibliographystyle{IEEEtran}
\bibliography{main}

\end{document}